\documentclass{emulateapj-rtx4}
\usepackage{natbib}
\usepackage{graphicx}
\usepackage{epstopdf}
\usepackage{apjfonts}
\usepackage{color}

\newcommand{\oiii}{[O\,{\footnotesize III}] $\lambda$5007}
\newcommand{\oiv}{[O\,{\footnotesize IV}] $\lambda$25.89$\micron$}
\newcommand{\ir}{12$\micron$}

\begin{document}
\title{On the Anisotropy of Nuclei Mid-Infrared Radiation in Nearby Active Galactic Nuclei}

\author{Huan~Yang, JunXian~Wang and Teng~Liu} 

\affil{CAS Key Laboratory for Research in Galaxies and Cosmology, Astronomy Department, University of Science and Technology of China, Hefei, Anhui 230026, China; yanghuan@mail.ustc.edu.cn; jxw@ustc.edu.cn; liuteng@ustc.edu.cn}

\begin{abstract}
In the center of active galactic nuclei (AGNs), the dusty torus absorb the radiation from the central engine and re-emit in mid-infrared (MIR). 
Observations have detected moderate anisotropy in the dust MIR emission, in the way that type 1 AGNs (type1s) are mildly brighter in MIR comparing with type 2 sources (type2s).
However, type1s and type2s were found to follow statistically the same tight MIR -- hard X-ray correlation, suggesting the MIR emission is highly isotropic assuming the hard X-ray radiation is inclination independent.
We argue this discrepancy could be solved considering the hard X-ray emission in AGN is also mildly anisotropic as we recently discovered.
To verify this diagram, we compare the sub-arcsecond \ir\ flux densities of type1s and type2s using \oiv\ emission line as an isotropic luminosity indicator.  
We find that on average type1s are brighter in nuclei \ir\ radiation by a factor of $2.6 \pm 0.6$ than type2s at given \oiv\ luminosities, confirming the mild anisotropy of the nuclei \ir\ emission. 
We show that the anisotropy of the \ir\ emission we detected is in good agreement with radiative transfer models of clumpy torus. 
The fact that type 1 and type 2 AGNs follow the same tight MIR -- hard X-ray correlation instead supports that both the MIR and hard X-ray emission in AGNs are mildly anisotropic.
\end{abstract}


\section{Introduction}

In the central a few tens parsecs of active galactic nuclei (AGN), the supermassive black holes are fueled by a lot of gas and dust 
~\citep[see review by][]{Alexander2012}.
The dusty clouds obscure a significant fraction of the sky of the central engines, and re-emit the absorbed energy mainly in mid-infrared (MIR). 
MIR observation therefore provides a powerful approach to infer the currently unclear geometric and physical properties of the dusty structure.  
Due to the angular momentum, the dusty gas likely forms a geometrically and optically thick torus.
Popular models include the so-called homogeneous torus model \citep[e.g.][]{Pier1992, Granato1997},
and the clumpy torus model \citep[e.g.][]{Krolik1988, Nenkova2008a, Nenkova2008, Honig2010}.

According to the unification scheme \citep{Antonucci1993}, type 1 and type 2 AGNs (hereafter type1s and type2s) are believed viewed along different inclination angles respect to the torus. 
As various torus models predict different levels of anisotropy in MIR emission from the torus, comparing the MIR radiation in type1s and type2s therefore could distinguish these models. 
It turns out that the homogenous torus model is disfavored since 
the MIR spectral energy distributions (SEDs) of type1s and type2s generally
show only mild difference, much weaker than expected from the homogenous torus model ~\citep[e.g.][]{Alonso2003, Hao2007, Wu2009, Tommasin2010, Ramos2011}.
Furthermore, these MIR spectra and SEDs are well fitted by the clumpy torus models which instead predict low to moderate level of anisotropy in MIR emission \citep[][]{Mason2006, Nenkova2008, Mor2009, Thompson2009, Nikutta2009, Ramos2009, Honig2010, Alonso2011, Lira2013}.

Quantifying the anisotropy in MIR observationally is essential to constrain the physical parameters of the torus models \citep[e.g.][]{Levenson2009, Honig2010a}. 
\citet{Heckman1995} reported that the average ratio of 10.6 $\micron$ to 1.4 GHz radio flux densities in type1s is larger by a factor of $\sim$ 4.0 than in type2s.
Similarly, \citet{Buchanan2006} explored the MIR anisotropy in Seyfert galaxies with Spitzer spectra after normalizing to 8.4 GHz radio emission. They found type1s are brighter than type2s by a factor of $\sim$ 7 at 5 $\micron$, and the factor of the difference gradually drops to $\sim$ 2 at 35 $\micron$.  
These studies suggest mild anisotropy of the MIR radiation in Seyfert galaxies, as the optical thin radio emission is believed an isotropic indicator of the intrinsic AGNs accretion power~\citep[e.g.][]{Diamond-Stanic2009}.

However, type 1 and type 2 AGNs appear to follow statistically the same tight MIR -- X-ray correlation  ~\citep{Lutz2004, Horst2006, Gandhi2009, Levenson2009, Honig2010a, Asmus2013}, suggesting the MIR emission is highly isotropic assuming the hard X-ray emission in AGNs is inclination independent. 
In this case, only tight upper limit to the anisotropy of MIR emission can be obtained, even with a rather large AGN sample containing 155 sources ~\citep[][]{Asmus2013}, contrary to the studies above which show that MIR emission is mildly anisotropic.

We argue this discrepancy is due to that the hard X-ray emission in AGNs is also mildly anisotropic. 
Adopting \oiv\ emission line as an isotropic AGN luminosity indicator~\citep[][]{Diamond-Stanic2009, Melendez2008, Rigby2009, Liu2010}, \citet[][hereafter paper I]{Liu2014} found both absorption corrected 2 -- 10 keV and SWIFT BAT 14 -- 195 keV emission in Compton-thin type2s are weaker than in type1s by a factor of 2 $\sim$ 3. This indicates the hard X-ray radiation in radio quiet AGNs is anisotropic, likely due to the beaming effect of an outflowing X-ray emitting corona.
Therefore, hard X-ray emission is likely not a good option to quantifying the moderate anisotropy in MIR emission in AGNs. In this work for the first time we utilize the \oiv\ emission line as the intrinsic luminosity proxy to measure the anisotropy of MIR \ir\ emission in nearby AGNs.

\section{The Sample}

\begin{figure*}[ht]
\centering
  \includegraphics[width=\textwidth]{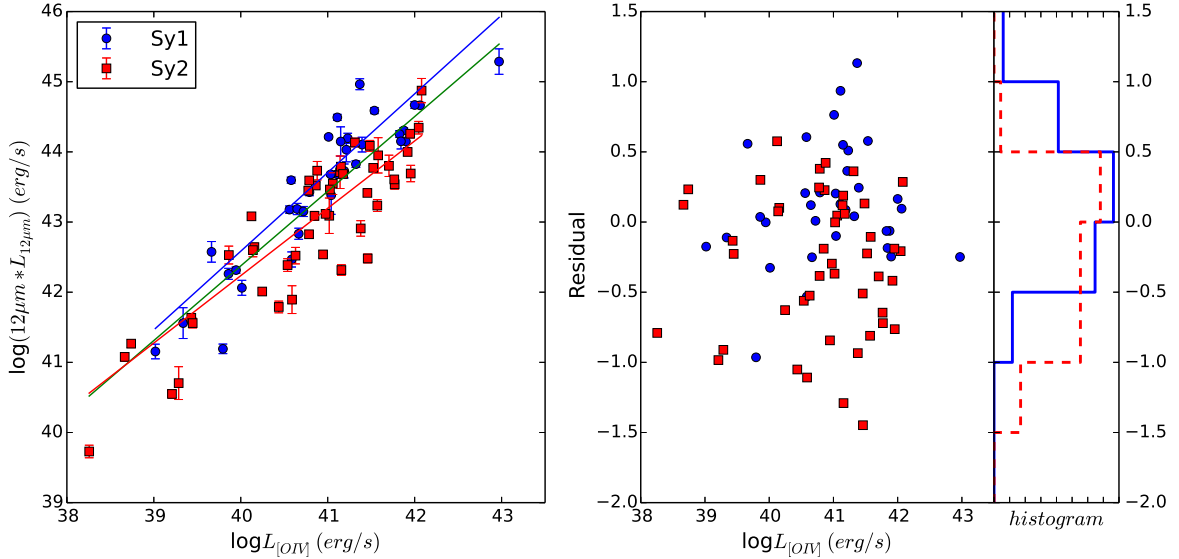}
 \caption{Left panel: The relations between \oiv\ and \ir\ luminosities. Blue dots are type 1 AGNs and red squares are type2s. The green, blue and red line are best linear fit lines for type1s + type2s, type1s and type2s respectively. Right panel: Residuals of \ir\ luminosity data points to the best-fit line of type1s + type2s. The histograms show the distributions of residuals for type1s (blue) and type2s (red) AGNs respectively. Type 2 AGNs are weaker in \ir\ luminosities than type1s.}
\end{figure*}

In Paper I, we have compiled a large sample of radio quiet AGNs with \oiv\ line fluxes measured by Spitzer Infrared Spectrometer from five major samples in literature \citep[][]{Diamond-Stanic2009, Weaver2010, Tommasin2010, Dasyra2009, Pereira-Santaella2010}.  
We match the \oiv\ catalog to the largest local AGN MIR photometry catalog \citep{Asmus2014}. \citet{Asmus2014} used all available sub-arcsecond resolution MIR imaging and measured the unresolved nuclear fluxes instead of aperture fluxes to minimize the contribution of possible extended emission.
Since our scientific goal is to study the MIR emission in typical AGNs,  all LINERs in the sample were dropped. 
As the jet in radio loud AGNs can contribute to MIR radiation, we also excluded known radio loud sources. 
The final sample includes 34 type 1 AGNs (including Seyfert 1, Seyfert 1.2 and Seyfert 1.5 galaxies, and a couple of type 1 quasars) and 47 type 2 AGNs (including Seyfert 1.8, Seyfert 1.9 and Seyfert 2) that have both \oiv\ and \ir\ luminosities. 
We note that dropping the quasars from our sample does not alter the results presented in this work.

The typical resolution of the \ir\ imaging observations is 0.3\arcsec\ \citep{Asmus2014}.
As the mean distance in log space is 82 (38) Mpc for type1s (type2s), with 1$\sigma$ scatter 0.5 (0.4) dex, the mean size of region probed by \ir\ imaging is 119 (56) parsecs for type1s (type2s).
At region scales about 50-100 parsecs, MIR spectroscopy observations of nearby AGNs show strongly reduced or undetected polycyclic aromatic hydrocarbon (PAH) emission compared to spectroscopy in a few hundreds parsecs to kiloparsecs scales \citep{Gandhi2009, Honig2010a, Asmus2014, Esquej2014}.  
As PAH emission is an indicator of the nuclear star formation activities, weak PAH emission suggests star formation has  limited contribution to \ir\ emission. 
Therefore the \ir\ emission mainly comes from dust in the center of Seyfert galaxies. 
The final matched sample is shown in Table 1. 
We note that this sample is neither complete nor homogeneous. 
Possible bias which may affect the results in this study is discussed in \S 3.

\section{Statistical Results}
We first plot the  \ir\ versus \oiv\ luminosity in Fig. 1.
We see that the \ir\ radiation strongly correlates to the \oiv\ for both type1s and type2s. 
The significance of the correlation, according to the Spearman Rank coefficient $\rho$, is $\rho\ = 0.85$ at a null significance level of $1 \times 10^{-23}$.
An intrinsic flux-flux correlation is also present, with $\rho\ = 0.61$ at a null significance level of $1 \times 10^{-9}$.

However, the \ir\ emissions at given \oiv\ luminosities are remarkably weaker in type2s than in type1s. 
To statistically quantifying the difference in the \ir\ luminosities between type1s and type2s, 
we first perform a linear fit to the relations on the total sample and subsamples of type1s and type2s,  taking \oiv\ as the independent variable. 
We only consider \ir\ luminosity uncertainties in the analysis, as only part of the sample have \oiv\ luminosities errors which are rather small (typical error is 5\%). 
The best-fitted lines are:
\[type1s + type2s: y = (1.06 \pm  0.07) x - 0.2 \pm 3.0\]
\[type1s: y = (1.12 \pm 0.11) x - 2.3 \pm 4.5\]
\[type2s: y = (0.96 \pm 0.08) x + 3.8 \pm 3.3\]
We then perform K-S test on the residuals (data points of \ir\ luminosities minus the best-fit line of total sample, see Fig. 1) of type1s and type2s.
At a confidence level of 99.8\%, they are not drawn from the same population .
The difference of the average residuals between type1s and type2s is 0.4 dex, thus type1s are brighter in \ir\  (at given \oiv\ luminosities) by a factor of $2.6 \pm 0.6$ than type2s, suggesting mild anisotropy in MIR radiation.  

We note that the statistical results on the difference between type1s and type2s are not sensitive to the linear-fit approach we utilized.
For instance, the orthogonal distance regression yields slightly different correlation slopes in \ir\ versus \oiv, but the resulted difference in the relative strength of \ir\ emission between type1s and type2s  remains unchanged, i.e., type1s are brighter in \ir\  by a factor of $2.6 \pm 0.6$ than type2s at a confidence level of 99.9\%.

Before looking into the underlying physics, we examine whether this difference could be artificial due to possible bias of the analysis. 
We first consider bias of the \oiv\ sample collection. 
The sources in the matched sample all have high resolution Spitzer IRS  spectra.
As type2s in the sample tend to have smaller redshifts/distances than type1s, possible slit loss might have 
underestimated the \oiv\ fluxes more severely in type2s comparing with type1s, correcting which would instead strengthen the difference we have detected. We refer the readers to \S4.2 in Paper I for more and thorough discussions on possible bias and the isotropy of \oiv. 

During the analysis, we have excluded 3 sources (all are type2s) with \oiv\ upper limits and 5  sources (all are type2s) with \ir\ upper limits. 
Including these upper limits with survival analysis yields consistent results, i.e., type1s are brighter in \ir\  by a factor of 2.8 than type2s at a confidence level of 99.95\%. 
We note there are 7 type 1.8 and type 1.9 AGNs in our sample, for which optical
classifications may be not robust. 
Excluding these 7 sources from the sample does not change the results in this work.
Simply classifying them as type 1 AGNs yield statistically the same results.

Could the difference between type1s and type2s shown in Fig. 1 be due to star formation contribution to \oiv\ as some works claimed that Seyfert 2 may have higher star
formation in their host galaxies (Maiolino et al. 1995; Buchanan et al. 2006; but see Imanishi \& Wada 2004)?
Comparing with \oiii, \oiv\ emission line has higher ionization potential (54.9 eV) thus is less affected by contamination from star formation in the host galaxy. 
The contamination to \oiv\ from star formation in the host galaxies is rather weak \citep{Pereira-Santaella2010} and the \oiv\ in Seyfert galaxies and quasars is generally dominated by the AGN. As shown in \citet{Pereira-Santaella2010} the star formation contribution to \oiv\ is much weaker at higher \oiv\ luminosity and could be significant at 
L{\scriptsize[OIV]} $<$ 10$^{40.2}$ erg~s$^{-1}$. 
The fraction of AGNs with L{\scriptsize[OIV]} $<$ 10$^{40.2}$ erg~s$^{-1}$ is 22\%.
Dropping those sources with L{\scriptsize[OIV]} $<$ 10$^{40.2}$ erg~s$^{-1}$ give statistically similar results: type1s are brighter in \ir\  by a factor of $3.5 \pm 0.9$ than type2s at a confidence level of 99.96\%.
Therefore the result in work can not be explained by possible contamination to \oiv\ from the host galaxies.

Finally, if there is any potential bias which could yield a spurious difference in the relative strength of \ir\  between type1s and type2s as we observed, we should have missed type1s with relatively weaker \ir\ emission, and/or type2s with relatively weaker \oiv\ emission.
Bias against sources with relatively brighter emission in one band is unlikely as astronomical observations are generally flux limited.
It is however rather puzzling if any such bias works preferentially on type1s but not type2s, or vice versa.
We conclude that although the sample is not complete or homogeneous, our major results in this work are not affected.

\section{Discussion}
\subsection{Both MIR and X-ray radiation Are Mildly Anisotropic}

The mild anisotropy in MIR radiation, as detected in previous studies \citep{Heckman1995, Buchanan2006}, is confirmed by this work. 
Particularly, the level of the anisotropy at \ir\ we detected (by a factor of 2.6) is in good agreement with Fig. 16 in \citet{Buchanan2006} at the same wavelength, which was however based on Spitzer spectra with lower spatial resolution comparing with this work.

Given that the X-ray emission is also mildly anisotropic (Paper I), the mild anisotropy in MIR radiation we detected is also consistent with the picture that type1s and type2s follow the same tight correlation between MIR and X-ray intrinsic luminosities ~\citep{Lutz2004, Horst2006, Gandhi2009, Levenson2009, Honig2010a, Asmus2013}. 

\subsection{Comparing with Clumpy Torus Models}

We quantitatively compare our results with two popular radiative transfer modelings of clumpy torus: CLUMPY ~\citep{Nenkova2008a, Nenkova2008} and CAT3D ~\citep{Honig2010}. 
Both models assume thermal equilibrium of dust with radiation of AGNs.
The models compute the dust radiation with a range of model parameters (Table 1),
including the optical opacity $\tau_{V}$ in V band of each cloud, 
the torus outer radius Y (Y = $R_{out}/R_{in}$, where $R_{in}$ is dust sublimation radius), the average number of clouds along the radial equatorial line N$_0$, the power-law index of clouds radial distribution  q , and half covering angle $\theta_0$ (see ~\citealp{Nenkova2008} and \citealp{Honig2010} for more details on the parameters).
Both models considered a Gaussian angular distribution of the clouds  $N \sim N_0$exp($-\beta^2/\theta_0^2)$, where
$\theta_0$ is the half covering angle, and $\beta$ is the angle of the line of sight from the equatorial plane. 
Both models generate SED grids for different sets of those parameters. Using these SED grids, we explore in which region in the parameter space the torus \ir\ radiation is mildly anisotropic as we observed.

\begin{deluxetable}{cccc}
\centering
\tablecaption{Parameters of Clumpy Torus Models We Adopted for Comparison}

\tablehead{\colhead{Parameters} & \colhead{Symbol} & \colhead{CLUMPY} & \colhead{CAT3D} }

\startdata
Opacity per single cloud in V band & $\tau_{V}$ & 60  & 50, 80 \\
Torus outer radius $R_{out}/R_{in}$ & Y & 100 & 150 \\
Clouds along radial equatorial rays & N$_0$ & [2, 15]  & [2.5, 10]\\
Clouds radial distribution $r^{-q}$  & q & [-3, 0] & [-2, 0] \\
Half covering angle & $\theta_0$ & [15$\arcdeg$, 60$\arcdeg$] &   45$\arcdeg$
\enddata
\end{deluxetable}

\begin{figure*}[ht]
\centering
  \includegraphics[width=\textwidth]{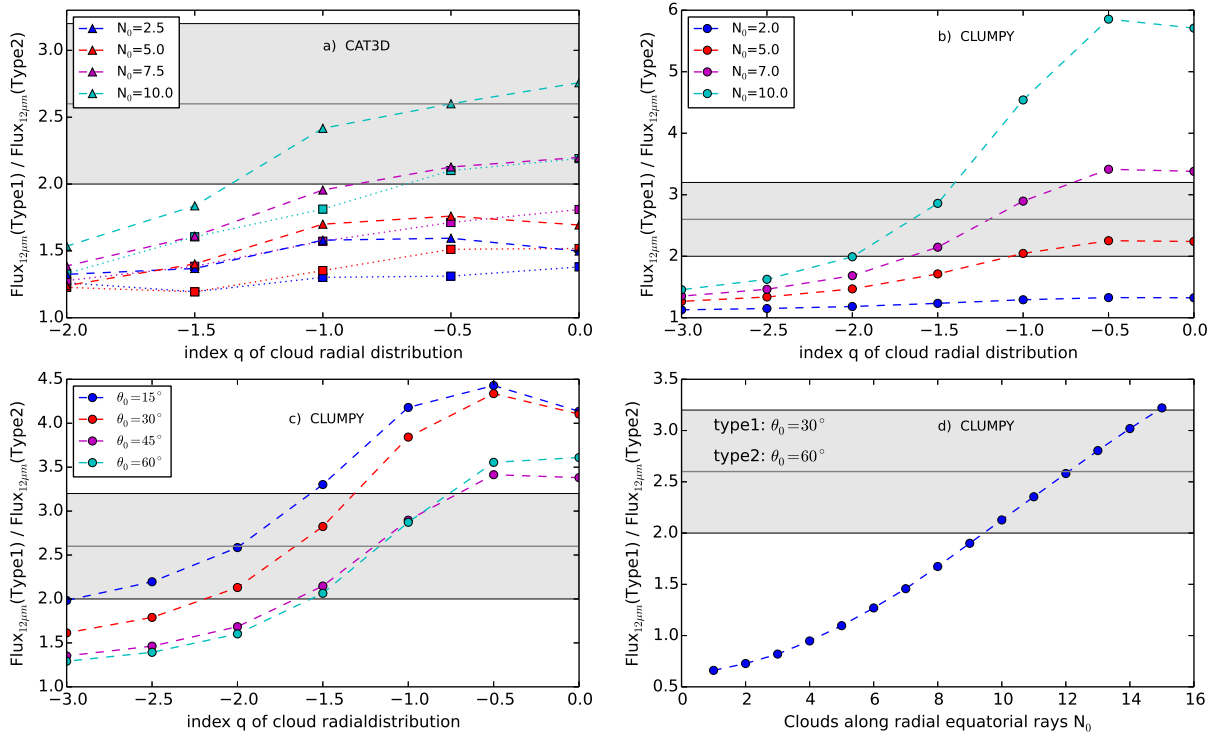}
 \caption{a) Flux ratio at \ir\ of type1s and type2s in clumpy torus model CAT3D for different model parameters. 
The torus half covering angle $\theta_0$ is fixed at 45$\arcdeg$. 
The dashed line and the triangles show results with $\tau_{V}$=80. The dotted line and the squares show results with $\tau_{V}$=50. 
b) same to a, but using CLUMPY model. c) same to b, but adopting different $\theta_0$ and fixed N$_0$ = 7. 
d) Assuming different $\theta_0$ for type1s and type2s, but the same viewing angle 50$\arcdeg$.  
$q$ is set to -1.0. 
The grey region plot the observed anisotropy of type1s and type2s.}
\end{figure*}

First, to compare emission of torus with different q and N$_0$, we take SED grids with parameters set to the model best-fit results of MIR spectra: $\theta_0$ = $45\arcdeg$, $\tau_{V}$ = 60 (50, 80), and Y = 100 (150) in CLUMPY (CAT3D) model \citep[][]{Nenkova2008, Thompson2009, Honig2010, Lira2013}, and explore how N$_0$ and q influence the anisotropy of \ir\ radiation. 
In clumpy torus models, the classification as type 1 or type 2 is probabilistic. 
We use the escape probability defined in \citet{Nenkova2008}  to calculate the average inclination angles of type 1 and type 2 AGNs (escape probability is defined as $\exp(-N_{LOS})$, while $N_{LOS}$ is the cloud number at the viewing line of sight),
For instance, adopting $\theta_0$ = $45\arcdeg$ and N$_0$=5.0 (the escape probability is 
$\exp(-5.0\exp((\frac{90\arcdeg-i}{45\arcdeg})^{2}))$ where i is the inclination angle), we obtained an average inclination angle of $30\arcdeg$ and $60\arcdeg$ respectively for type 1 and type 2 AGNs (the escape probabilities at the average inclination angles are 18.4\% and 0.1\% for type1s and type2s).
We roughly estimate the anisotropy as the ratio of \ir\ flux viewed at the average inclination angles of type1s and type2s.
As shown in Fig. 2, the anisotropy increases when the index of cloud radial distribution q is flatter and the clouds numbers N is larger, consistent with Figure 10 in \citet{Nenkova2008}. 

We then explore how the parameter half covering angle $\theta_0$ of the torus affects the anisotropy in dust emission.
We adopt CLUMPY SED model as it provides more grids of the physical parameters. 
We set N$_0$ = 7, $\tau_{V}$ = 60, and Y = 100. 
We compute the average inclination angles of type1s and type2s for different $\theta_0$ and estimate the anisotropy as the ratio of \ir\ flux viewed at the respective average inclination angle. 
As shown in Fig. 2c, the anisotropy increases when the half covering angle $\theta_0$ is smaller.

By fitting a sample of AGN SEDs with clumpy torus model, an interesting finding \citep[][]{Ramos2011} suggests type 1 and type 2 AGNs have a statistical difference in the half covering angle of the torus, but similar probability distributions of the viewing angles.
In this case, we assume different half covering angles for two types ($\theta_0$ = 30$\arcdeg$ for type1s and $\theta_0$ = 60$\arcdeg$ for type2s), but the same average viewing angle of 50$\arcdeg$.
We set $q$ = -1.0.
The result shows that in this scheme, the observed anisotropy could also be reproduced with N$_0$ = 10 -- 15.
Note that with much smaller $N_0$, we would expect opposite anisotropy that type2s are brighter in MIR than type1s, due to its larger covering angle of the dusty torus.

As there is degeneracy among the torus parameters which affect the anisotropy, in this work we are unable to give accurate constraints to individual parameters. 
The anisotropy of a factor of $\sim$ 2.6 in dust MIR \ir\ radiation requires a large N$_0$ and/or a flat q and/or a small $\theta_0$.
Those parameter ranges are generally consistent with the best-fit results to MIR spectra \citep[][]{Thompson2009, Honig2010, Lira2013}, particularly \citet{Alonso2011} and \citet{Ramos2011} which include both type 1 and type 2 Seyferts.  
We conclude that the mild anisotropy we observed in MIR emission is fully consistent with modern clumpy torus models.

\subsection{Considering Dust Emission From The Polar Region}

Although it is a popular picture that the dust MIR emission in AGNs is dominated by radiation from the clumpy torus, it is challenged by recent MIR interferometric observations. 
Infrared interferometry observations of 4 nearby AGNs (Circinus, NGC1068, NGC424, NGC3783) at sub-pc resolution suggested that, instead of the torus, the major contributor of MIR emission is dust in the polar region ~\citep{Tristram2007, Raban2009, Honig2012, Honig2013, Tristram2014}. 
Assuming all AGNs contain polar region dust, the dust emission in polar region will likely dilute the anisotropic emission from the torus. 

The level of the dilution is however hard to estimate. 
The anisotropy of total \ir\ emission depends on the contribution from both the torus and the polar region. 
The polar region MIR emission is also likely different between type1s and type2s as the sky coverage of the polar region is likely larger in type1s due to selection effect (Sources with smaller sky coverage of the dusty torus thus larger sky coverage of the polar region are more likely detected as type1s.).
This effect may have played a role in producing the observed MIR anisotropy in AGNs if the polar region component dominate the MIR emission.
We note that if type1s have more  \ir\ emission from torus (relative to that from the polar region) than type2s, i.e, the anisotropy in the MIR emission is dominated by the torus but not the polar region, the MIR emission from type1s should appear more compact. 
Recently, the MIR interferometry observations of a sample of AGNs showed that type1s have a larger fraction of emission concentrated in the central unresolved point sources ~\citep{Burtscher2013}, which seems consistent with this diagram.

\acknowledgments
We thank Daniel Asmus for helpful discussion and comments, and the anonymous referee for valuable suggestions that helped improve the manuscript.
This work is supported by Chinese NSF through grant 11233002 \& 11421303.
JXW acknowledges support from Chinese Top-notch Young Talents Program and National Basic Research Program of China (973 program, grant No. 2015CB857005).

\clearpage
\LongTables

\begin{deluxetable}{ccccccccc}
\centering
\tablecaption{The Sample}
\tablewidth{15pt}

\tablehead{\colhead{Name} & \colhead{RA} & \colhead{DEC} & \colhead{Redshift} & \colhead{D} & \colhead{Type} & \colhead{$\log L_{[OIV]}$} & \colhead{$\log L_\nu(12\micron)$} & \colhead{error of $\log L_\nu(12\micron)$} \\
\colhead{} & \colhead{(deg.)} & \colhead{(deg.)} & \colhead{} & \colhead{(Mpc)} & \colhead{} & \colhead{($erg~s^{-1}$)} & \colhead{($erg~s^{-1}$)} & \colhead{(dex)} } 

\startdata
PG 0026+129 & 7.30667 & 13.26750 & 0.1420 & 691.0 & 1.2 & 42.06 & 44.66 & 0.04 \\
I Zw1 & 13.39558 & 12.69339 & 0.0589 & 269.0 & 1 & 41.37 & 44.96 & 0.08 \\
NGC 424 & 17.86512 & -38.08347 & 0.0118 & 49.5 & 2 & 40.88 & 43.73 & 0.13 \\
Fairall 9 & 20.94075 & -58.80578 & 0.0470 & 215.0 & 1.2 & 41.54 & 44.59 & 0.04 \\
NGC 526A & 20.97662 & -35.06553 & 0.0191 & 82.8 & 1.9 & 41.18 & 43.69 & 0.05 \\
Mrk 1014 & 29.95921 & 0.39461 & 0.1631 & 807.0 & 1.5 & 42.97 & 45.29 & 0.18 \\
NGC 788 & 30.27687 & -6.81553 & 0.0136 & 57.2 & 2 & 40.98 & 43.12 & 0.05 \\
Mrk 590 & 33.63983 & -0.76669 & 0.0264 & 116.0 & 1 & 40.58 & 43.60 & 0.04 \\
NGC 1068 & 40.66963 & -0.01328 & 0.0038 & 14.4 & 2 & 41.70 & 43.80 & 0.15 \\
NGC 1144 & 43.80083 & -0.18356 & 0.0288 & 128.0 & 2 & 41.02 & 43.09 & 0.25 \\
MCG-2-8-39 & 45.12746 & -11.41572 & 0.0299 & 133.0 & 2 & 41.48 & 44.09 & 0.08 \\
NGC 1194 & 45.95462 & -1.10375 & 0.0136 & 58.2 & 1.9 & 40.78 & 43.45 & 0.04 \\
NGC 1365 & 53.40154 & -36.14039 & 0.0055 & 17.9 & 1.8 & 40.95 & 42.54 & 0.04 \\
NGC 1386 & 54.19242 & -35.99942 & 0.0029 & 16.5 & 2 & 40.54 & 42.39 & 0.09 \\
NGC 1566 & 65.00175 & -54.93781 & 0.0050 & 14.3 & 1.5 & 39.34 & 41.56 & 0.22 \\
NGC 1667 & 72.15475 & -6.31997 & 0.0152 & 67.8 & 2 & 40.59 & 41.89 & 0.20 \\
MCG-1-13-25 & 72.92283 & -3.80925 & 0.0159 & 71.2 & 1.2 & 39.66 & 42.57 & 0.15 \\
ESO 33-2 & 73.99567 & -75.54117 & 0.0181 & 82.3 & 2 & 41.06 & 43.55 & 0.16 \\
Ark 120 & 79.04758 & -0.14983 & 0.0327 & 149.0 & 1 & 41.01 & 44.21 & 0.02 \\
ESO 362-18 & 79.89917 & -32.65758 & 0.0124 & 56.5 & 1.5 & 40.56 & 43.18 & 0.05 \\
IRAS 05189-2524 & 80.25579 & -25.36261 & 0.0426 & 196.0 & 2 & 42.08 & 44.87 & 0.17 \\
ESO 253-3 & 81.32533 & -46.00583 & 0.0425 & 196.0 & 2 & 42.05 & 44.34 & 0.09 \\
NGC 2110 & 88.04742 & -7.45622 & 0.0078 & 35.9 & 2 & 40.85 & 43.09 & 0.06 \\
H0557-385 & 89.50833 & -38.33464 & 0.0339 & 156.0 & 1.2 & 41.11 & 44.49 & 0.04 \\
ESO 5-4 & 91.42346 & -86.63186 & 0.0062 & 22.4 & 2 & 39.43 & 41.64 & 0.05 \\
Mrk 3 & 93.90150 & 71.03753 & 0.0135 & 60.6 & 2 & 41.95 & 43.69 & 0.12 \\
PG 0844+349 & 131.92696 & 34.75122 & 0.0640 & 302.0 & 1 & 41.21 & 44.03 & 0.17 \\
MCG-1-24-12 & 140.19271 & -8.05614 & 0.0196 & 93.8 & 2 & 41.03 & 43.46 & 0.04 \\
MCG-5-23-16 & 146.91733 & -30.94872 & 0.0085 & 42.8 & 1.9 & 40.79 & 43.59 & 0.04 \\
Mrk 1239 & 148.07958 & -1.61208 & 0.0199 & 95.4 & 1 & 41.23 & 44.19 & 0.07 \\
NGC 3081 & 149.87308 & -22.82628 & 0.0080 & 40.9 & 2 & 41.38 & 42.91 & 0.11 \\
NGC 3227 & 155.87742 & 19.86506 & 0.0039 & 22.1 & 1.5 & 40.59 & 42.47 & 0.11 \\
NGC 3281 & 157.96704 & -34.85369 & 0.0107 & 52.8 & 2 & 41.77 & 43.61 & 0.05 \\
NGC 3783 & 174.75733 & -37.73867 & 0.0097 & 48.4 & 1.5 & 41.03 & 43.68 & 0.03 \\
NGC 3982 & 179.11721 & 55.12525 & 0.0037 & 21.4 & 2 & 39.45 & 41.56 & 0.07 \\
NGC 4051 & 180.79004 & 44.53133 & 0.0023 & 12.2 & 1 & 39.95 & 42.31 & 0.04 \\
NGC 4138 & 182.37408 & 43.68531 & 0.0030 & 13.8 & 1.9 & 38.67 & 41.08 & 0.06 \\
NGC 4151 & 182.63575 & 39.40572 & 0.0033 & 13.3 & 1.5 & 40.67 & 42.83 & 0.08 \\
NGC 4235 & 184.29117 & 7.19158 & 0.0080 & 41.2 & 1.2 & 39.86 & 42.26 & 0.08 \\
NGC 4258 & 184.73958 & 47.30397 & 0.0015 & 7.6 & 2 & 38.74 & 41.27 & 0.06 \\
NGC 4388 & 186.44479 & 12.66208 & 0.0084 & 19.2 & 2 & 41.16 & 42.32 & 0.08 \\
NGC 4395 & 186.45358 & 33.54692 & 0.0011 & 4.3 & 1.8 & 38.26 & 39.73 & 0.09 \\
NGC 4501 & 187.99650 & 14.42039 & 0.0076 & 17.9 & 2 & 39.21 & 40.55 & 0.06 \\
NGC 4507 & 188.90263 & -39.90925 & 0.0118 & 57.5 & 2 & 41.15 & 43.79 & 0.04 \\
NGC 4593 & 189.91429 & -5.34425 & 0.0090 & 45.6 & 1 & 40.72 & 43.15 & 0.07 \\
IC 3639 & 190.22021 & -36.75586 & 0.0109 & 53.6 & 2 & 40.86 & 43.52 & 0.04 \\
NGC 4941 & 196.05475 & -5.55161 & 0.0037 & 21.2 & 2 & 40.25 & 42.01 & 0.05 \\
ESO 323-77 & 196.60887 & -40.41467 & 0.0150 & 71.8 & 1.2 & 41.19 & 43.73 & 0.10 \\
NGC 5033 & 198.36446 & 36.59394 & 0.0029 & 18.1 & 1.2 & 39.79 & 41.19 & 0.07 \\
MCG-3-34-64 & 200.60192 & -16.72847 & 0.0165 & 79.3 & 2 & 41.92 & 44.00 & 0.05 \\
NGC 5135 & 201.43358 & -29.83367 & 0.0137 & 66.0 & 2 & 41.57 & 43.24 & 0.08 \\
M51a & 202.46962 & 47.19517 & 0.0015 & 8.1 & 2 & 39.29 & 40.70 & 0.23 \\
MCG-6-30-15 & 203.97379 & -34.29553 & 0.0077 & 38.8 & 1.5 & 40.65 & 43.19 & 0.08 \\
NGC 5273 & 205.53475 & 35.65422 & 0.0035 & 15.3 & 1.5 & 39.02 & 41.15 & 0.10 \\
IC 4329A & 207.33029 & -30.30944 & 0.0161 & 76.5 & 1.2 & 41.87 & 44.31 & 0.04 \\
NGC 5347 & 208.32429 & 33.49083 & 0.0078 & 38.1 & 2 & 40.12 & 43.08 & 0.04 \\
Circinus & 213.29146 & -65.33922 & 0.0014 & 4.2 & 2 & 40.16 & 42.64 & 0.05 \\
NGC 5506 & 213.31204 & -3.20758 & 0.0062 & 31.6 & 2 & 41.46 & 43.41 & 0.03 \\
NGC 5548 & 214.49804 & 25.13678 & 0.0172 & 80.7 & 1.5 & 41.04 & 43.38 & 0.27 \\
NGC 5643 & 218.16975 & -44.17442 & 0.0040 & 20.9 & 2 & 40.63 & 42.52 & 0.12 \\
NGC 5728 & 220.59958 & -17.25308 & 0.0094 & 45.4 & 2 & 41.46 & 42.48 & 0.06 \\
IC 4518W & 224.42158 & -43.13211 & 0.0163 & 76.1 & 2 & 41.77 & 43.54 & 0.07 \\
Mrk 841 & 226.00500 & 10.43783 & 0.0364 & 170.0 & 1.5 & 41.90 & 44.15 & 0.13 \\
NGC 5995 & 237.10396 & -13.75778 & 0.0252 & 117.0 & 1.9 & 41.31 & 44.13 & 0.06 \\
NGC 6300 & 259.24779 & -62.82056 & 0.0037 & 14.3 & 2 & 39.86 & 42.53 & 0.13 \\
Fairall 49 & 279.24288 & -59.40239 & 0.0200 & 90.1 & 2 & 41.58 & 43.95 & 0.25 \\
ESO 103-35 & 279.58475 & -65.42756 & 0.0133 & 59.5 & 2 & 41.14 & 43.71 & 0.23 \\
Fairall 51 & 281.22492 & -62.36483 & 0.0142 & 64.1 & 1.5 & 41.11 & 43.68 & 0.04 \\
ESO 141-55 & 290.30892 & -58.67031 & 0.0371 & 169.0 & 1.2 & 41.39 & 44.10 & 0.10 \\
NGC 6814 & 295.66933 & -10.32350 & 0.0052 & 20.1 & 1.5 & 40.01 & 42.06 & 0.11 \\
NGC 6860 & 302.19537 & -61.10019 & 0.0149 & 65.8 & 1.5 & 40.79 & 43.43 & 0.05 \\
NGC 6890 & 304.57542 & -44.80672 & 0.0081 & 33.8 & 2 & 40.14 & 42.60 & 0.10 \\
Mrk 509 & 311.04058 & -10.72347 & 0.0344 & 153.0 & 1.5 & 41.83 & 44.25 & 0.05 \\
IC 5063 & 313.00975 & -57.06878 & 0.0113 & 49.1 & 2 & 41.52 & 43.77 & 0.03 \\
PG 2130+099 & 323.11588 & 10.13875 & 0.0630 & 288.0 & 1.5 & 42.00 & 44.67 & 0.05 \\
NGC 7172 & 330.50787 & -31.86967 & 0.0087 & 34.8 & 2 & 40.78 & 42.83 & 0.04 \\
Mrk 304 & 334.30108 & 14.23914 & 0.0658 & 301.0 & 1 & 41.15 & 44.15 & 0.21 \\
NGC 7314 & 338.94246 & -26.05047 & 0.0048 & 18.3 & 2 & 40.44 & 41.79 & 0.08 \\
NGC 7469 & 345.81508 & 8.87400 & 0.0163 & 67.9 & 1.5 & 41.32 & 43.83 & 0.05 \\
Mrk 926 & 346.18117 & -8.68572 & 0.0469 & 210.0 & 1.5 & 41.84 & 44.15 & 0.11 \\
NGC 7674 & 351.98633 & 8.77903 & 0.0289 & 126.0 & 2 & 41.95 & 44.26 & 0.06 \\

\enddata

\tablecomments{(1) Commonly used object names; (2-4) Ra, Dec, redshifts from NED; (5) Distances from Asmus et al. 2014, most of which are redshift-based luminosity distances with $H_{0}$ = 67.3, $\Omega_{m}$ = 0.315, and $\Omega_{vac}$ = 0.685 ~\citep{Planck2013}, plus a few redshift-independent distances; (6) Optical AGN classifications from paper I; (7) Luminosities of [O\,{\footnotesize IV}] $\lambda$25.89$\micron$ from paper I; 
(8-9) Nuclear sub-arcsecond scale monochromatic luminosities and errors at rest frame 12$\micron$~\citep{Asmus2014}.}
\end{deluxetable}

\clearpage

\end{document}